\documentclass[pdflatex,sn-basic]{sn-jnl}

\usepackage{pdflscape}
\usepackage{caption}
\usepackage{subcaption}
\usepackage{paralist}
\usepackage{siunitx}
\usepackage{chngcntr} 
\usepackage{graphicx}%
\usepackage{multirow}%
\usepackage{amsmath,amssymb,amsfonts}%
\usepackage{amsthm}%
\usepackage{mathrsfs}%
\usepackage[title]{appendix}%
\usepackage{xcolor}%
\usepackage{textcomp}%
\usepackage{manyfoot}%
\usepackage{booktabs}%
\usepackage{algorithm}%
\usepackage{algorithmicx}%
\usepackage{algpseudocode}%
\usepackage{listings}%

\theoremstyle{thmstyleone}%
\theoremstyle{thmstyletwo}%

\theoremstyle{thmstylethree}%
\newtheorem{definition}{Definition}%
\newtheorem{problem}{Problem}%

\makeatletter
\renewcommand{\@thefnmark}{} 
\@ifundefined{affil}{\newcommand{\affil}[2][]{#2}}{}
\@ifundefined{fnm}{}{}
\@ifundefined{sur}{}{}
\@ifundefined{orgdiv}{}{}
\@ifundefined{orgname}{}{}
\@ifundefined{orgaddress}{}{}
\@ifundefined{city}{}{}
\@ifundefined{country}{}{}
\makeatother

\begin{document}


\title{MOUFLON: Multi-group Modularity-based Fairness-aware Community Detection}
\author{
Georgios Panayiotou\thanks{georgios.panayiotou@it.uu.se},\,
Anand Mathew Muthukulam Simon,\,
Matteo Magnani,\,
Ece Calikus
}

\affil{InfoLab, Department of Information Technology, Uppsala University, Uppsala, Sweden}

\abstract{
In this paper, we propose MOUFLON, a fairness-aware, modularity-based community detection method that allows adjusting the importance of partition quality over fairness outcomes. 
MOUFLON uses a novel proportional balance fairness metric, providing consistent and comparable fairness scores across multi-group and imbalanced network settings.  
We evaluate our method under both synthetic and real network datasets, focusing on performance and the trade-off between modularity and fairness in the resulting communities, along with the impact of network characteristics such as size, density, and group distribution. 
As structural biases can lead to strong alignment between demographic groups and network structure, we also examine scenarios with highly clustered homogeneous groups, to understand how such structures influence fairness outcomes.
Our findings showcase the effects of incorporating fairness constraints into modularity-based community detection, and highlight key considerations for designing and benchmarking fairness-aware social network analysis methods.
}

\maketitle

\section{Introduction}\label{sect:introduction}
Given the increasing use of automation in high-stakes decision-making systems, algorithmic fairness has emerged as a critical field of study, shifting attention beyond traditional performance metrics such as accuracy and execution time \citep{dwork_fairness_2012,pessach_algorithmic_2023,caton_fairness_2024}.
Concepts of algorithmic fairness have been discussed in the context of a variety of critical decision-making systems, such as pre-trial risk assessment \citep{mitchell_algorithmic_2021}, credit scoring \citep{das_algorithmic_2023}, and education \citep{kleinberg_algorithmic_2018}.
Recently, research in algorithmic fairness has expanded to also include social network analysis. Recent studies have shown that use of fairness-oblivious social network analysis methods can reinforce power asymmetries and demographic biases, due to structural properties such as homophily and class imbalances \citep{avin_modeling_2017,masrour_bursting_2020,espin-noboa_inequality_2022,sekara_detecting_2024,oliveira_stronger_2024}. As a result, algorithmic fairness in social network analysis has emerged as a nascent field of study, where the central goal is to design methods that mitigate network inequalities found in the data, ensuring outcomes do not disproportionately favor any particular groups \citep{saxena_fairsna_2024}.

With the use of network data to represent social networks in very large platforms, including online social media and online retail systems, fairness-aware community detection methods become essential for preventing biased outcomes in algorithmic decision-making. 
In such large-scale systems, community detection methods are often used to identify groups of similar users and provide tailored recommendations to users belonging to the same group \citep{eirinaki_recommender_2018,ying_preference-aware_2013}.
However, the application of fairness-oblivious community detection methods that rely solely on subgraph density, as seen in the majority of state-of-the-art methods, may have detrimental effects. For example, in online social media platforms such as Facebook and X, using community-based recommendations \citep{meta_ai_2023,twitter_twitters_2023}, may result in individuals with extremely similar interests and political views being grouped together, potentially leading to filter bubbles and reduced access to the diversity of opinions \citep{pariser_filter_2012}. Instead, we need methods to identify well-connected communities to ensure relevance of the provided services (e.g., recommendations), but at the same time maintain a diverse representation of demographics in the communities, so as to avoid bias stemming from disproportional representation.
Moreover, fairness-aware community detection can prove useful in studying influence spreading \citep{stoica_fairness_2019}, and when producing balanced communities for randomized experiments \citep{saveski_detecting_2017}.

Although fairness-aware community detection is beginning to attract attention, it unfortunately remains relatively underexplored \citep{dong_fairness_2023,saxena_fairsna_2024}. 
While there has been significant recent effort focusing on fair clustering of various data types \citep{bera_fair_2019,chhabra_overview_2021}, research has primarily focused on non-graph applications, thus ignoring the underlying network structure necessary to identify meaningful communities. 
In contrast, fairness-aware methods for graph clustering are still limited, with existing efforts primarily exploring spectral approaches \citep{kleindessner_guarantees_2019,wang_scalable_2023}, which both require a number of expected clusters as a parameter, and generally are computationally expensive due to eigenvalue decomposition, thus hindering scalability to large networks.

Modularity-based community detection, albeit efficient, scalable, and widely used in practical applications \citep{fortunato_community_2010,malliaros_clustering_2013,li_comprehensive_2024}, has only recently been targeted in the context of fairness-aware community detection. 
\citet{panayiotou_fair-mod_2025} introduced Fair-mod, a fairness-aware Louvain-based community detection method considering the well-studied fairness notion of group balance based on the doctrine of disparate impact.
Concurrently, \citet{gkartzios_fair_2025} proposed a modification of the Louvain algorithm by constraining node movement on modularity-based fairness metrics.
However, both approaches are limited to assessing demographic fairness between only two groups. In addition, the maximum attainable fairness scores can be greatly affected by network structure and demographic group imbalances. This effect can be especially relevant for modularity-based fairness metrics. Consequently, it becomes difficult to distinguish whether low fairness scores reflect unfairness toward a group or simply structural constraints of the network. 
This key gap motivates our work: a new fairness metric that (i) extends to settings with more than two demographic groups, and (ii) remains robust to possible class imbalances in the network.

Another crucial yet underexplored challenge is the inherent trade-off between community quality and fairness outcomes. While optimizing for modularity alone often leads to partitions that reinforce latent biases, enforcing strict fairness constraints can distort the underlying community structure. The need to address this tension between partition quality and fairness has been highlighted by various recent studies \citep{hakim_fairness-quality_2024, panayiotou_towards_2025, gkartzios_fair_2025}. Despite this, no fairness-aware community detection method exists that allows controlling the trade-off between partition quality and fairness, while simultaneously considering multiple sensitive demographic groups.

In this work, we aim to address the aforementioned challenges by integrating fairness constraints into modularity-based community detection with a scalar objective, allowing for adjusting the importance of partition quality over fairness outcomes. 
We propose MOUFLON, a multi-group fairness-aware method for modularity-based community detection, aiming to obtain communities that are both highly modular and balanced with respect to the distribution of demographic groups in the network. As part of our method, we also introduce a novel group fairness metric for proportional balance within a community. 

We experimentally evaluate our method, primarily focusing on scalability and the trade-off between modularity and fairness scores in the resulting communities. 
Since existing biases in the network can lead to strong alignment between demographics and network structure, we also study how different group distributions and network structures can affect the optimization process, an aspect often overlooked in previous work. Thus, we also examine our method under a variety of network and group distribution settings, including extreme scenarios where single demographic groups form strongly connected groups, as such structures might directly impact the types of partitions that can be identified, and in turn, their fairness outcomes. 

\vspace{2cm}
Our contributions can be summarized as follows: 
\begin{itemize}
    \item We introduce MOUFLON, a modification of the Louvain algorithm that optimizes a scalar objective of modularity and fairness.
    \item We propose a novel proportional balance fairness metric, allowing consistent fairness outcomes across multi-group and imbalanced network settings.
    \item We conduct an ablation study under various network and group distribution settings, focusing on performance, trade-off between modularity and fairness, types of communities identified by the method, and impact of strongly connected groups on fairness outcomes.
\end{itemize}

\section{Related work}\label{sect:related}
The field of algorithmic fairness in clustering has recently attracted substantial attention \citep{bera_fair_2019,chhabra_overview_2021}. A prominent example is the seminal work on fairlets by \citet{chierichetti_fair_2017}, codifying the notion of disparate impact \citep{feldman_certifying_2015}. Other relevant works include methods for fair correlation \citep{ahmadian_fair_2020}, hierarchical \citep{ahmadian_fair_2020-1} and probabilistic clustering \citep{esmaeili_probabilistic_2020}. However, these works focus on vector-based data, which do not capture the connectivity patterns and relational structures in networks, thus leading to community membership primarily dependent on feature similarity rather than network topology. 

Fair clustering has also been explored within graph settings, with most applications focusing on fairness-aware spectral \citep{kleindessner_guarantees_2019,wang_scalable_2023} and correlation clustering \citep{gullo_when_2022,casel_fair_2023}. While these approaches extend fairness-aware clustering to network data, they share the common challenge of scalability to large networks, especially with respect to space complexity. In addition, neither variant offers a mechanism to prioritize the importance of network structure over partition fairness.
Fair graph clustering methods primarily rely on two fairness classes that are also common in non-graph clustering \citep{chhabra_overview_2021}. Group-based fairness notions, also referred to as demographic- or pattern-based, require that groups under a sensitive attribute are proportionally represented in each cluster, typically through balance constraints \citep{dong_fairness_2023}. Such group fairness constraints have been applied in fairness-aware spectral \citep{kleindessner_guarantees_2019,wang_scalable_2023} and correlation clustering \citep{gullo_when_2022,casel_fair_2023}. In contrast, individual fairness does not focus on sensitive attributes, but instead requires that the neighbors of a node are proportionally distributed across clusters \citep{dong_fairness_2023,ghodsi_towards_2024}. In this work, we specifically focus on group fairness constraints for community detection; note that we distinguish community detection as a special case of graph clustering, where the goal is to find a partition into communities, forming densely connected subgraphs with sparse connections between them \citep{fortunato_community_2010}. 

Research on fairness in community detection is more scarce, as \citet{saxena_fairsna_2024} note, but has gained increasing attention, with recent works evaluating the fairness of conventional community detection algorithms.  
\citet{de_vink_group_2025} and \citet{de_vink_quantifying_2025} highlight the differences in partition fairness of various traditional, fairness-oblivious community detection methods. The authors also propose various community size-based fairness objectives; however, they do not consider attributed networks. In this work, we focus on networks with information about node demographics, so these would not be usable for our problem. 
\citet{manolis_modularity-based_2023} evaluate the partitions produced by the Louvain algorithm \citep{blondel_fast_2008} on social networks, and introduce a modularity-based group fairness definition. While neither of the aforementioned studies proposes fair community detection methods based on these metrics, they highlight how the use of fairness-oblivious community detection algorithms might produce low-fairness clusters, thus providing an important motivation for our work. 

The first fairness-aware community detection methods targeting demographic fairness objectives have been recently proposed by \citet{panayiotou_fair-mod_2025} and \citet{gkartzios_fair_2025}. 
\citet{panayiotou_fair-mod_2025}, proposed Fair-mod, a modification of the Louvain community detection algorithm, optimizing for a weighted sum over modularity and group balance in the obtained partition. 
\citet{gkartzios_fair_2025} proposed Fairness-Aware Louvain, a fairness-aware community detection method implementing various modularity-based fairness constraints into the Louvain community detection algorithm. Contrary to Fair-mod, Fairness-Aware Louvain constrains node movement between communities that would worsen fairness, without, however, proposing a method to offer a trade-off between partition quality and fairness. 
However, both methods are limited to two demographic groups, ignoring thus sensitive attributes containing multiple classes, and are not designed to distinguish whether low fairness scores reflect network structure and class imbalances.

In this work, we address these gaps by proposing MOUFLON, a scalable modularity-based community detection method, which uses a proportional balance-based fairness metric for multiple demographic groups. We refer to Table \ref{tab:fair-cd-comparison} for a summary of the properties of existing approaches for fair graph clustering and community detection. In addition, we also conduct an ablation study that examines how different factors, namely the choice of fairness score, the importance of community quality over fairness, and the proportion of groups in the network, affect fairness outcomes. We also consider often overlooked aspects of the network, such as the presence of clustered homogeneous groups, which can also affect the performance of our method by influencing the types of communities that can be discovered.

\begin{table}[tb!]
\centering
\resizebox{\textwidth}{!}{
\begin{tabular}{lccccc}
\toprule
\textbf{Method}                     & \textbf{Sc} & \textbf{AC} & \textbf{Q/F} & \textbf{MG}    & \textbf{Prop} \\
\midrule
Fair spectral graph clustering      &             &             &              & \checkmark     & \\
Fair correlation graph clustering   &             & \checkmark  &              & \checkmark     & \\
\midrule
Fairness-aware Louvain              & \checkmark  & \checkmark  &              &                & \\
Fair-mod                            & \checkmark  & \checkmark  & \checkmark   &                & \\
MOUFLON                             & \checkmark  & \checkmark  & \checkmark   & \checkmark     & \checkmark \\
\bottomrule
\end{tabular}
}
\caption{Supported features of fairness-aware graph clustering and community detection methods. We denote with a $\checkmark$ the methods that are scalable to large networks (Sc), are non-parametric to the number of communities (AC), can adjust to prioritize partition quality over fairness (Q/F), offer support for non-binary sensitive attributes with more than two groups (MG), and are based on fairness criteria with scores resistant to group proportions in the network (Prop)}
\label{tab:fair-cd-comparison}
\end{table}

\section{Fair modular community detection}\label{sect:preliminaries}
In this section, we introduce the notation, definitions, and concepts used throughout the paper. We first review the network model and modularity quality function for community detection, and then state our problem formulation.

\subsection{Network preliminaries}\label{ssec:network}
We consider an undirected, weighted network $G=(V,E)$ with $n=|V|$ nodes and $m=|E|$ edges. The adjacency matrix $A$ has entries $A_{ij}=A_{ji}=w_{ij}$ if $(i,j) \in E$ and $A_{ij}=A_{ji}=0$ otherwise, where $w_{ij} \geq 0$ is a non-negative weight. The weighted degree (or strength) of a node $i$ is given by $k_i=\sum_{j \in V} A_{ij}$.

\paragraph{Modularity}
One of the most commonly used quality functions for community detection is modularity \citep{newman_modularity_2006}, which compares the observed density of edges within communities to that expected under a random graph null model preserving node degrees. 
For a partition $P$ of the network's nodes $V$, the modularity score of the partition $Q_P$ is given by:
\begin{equation}\label{eq:modularity}
    Q_{P} = \frac{1}{2m} \sum_{C \in P} \sum_{i,j \in C} (A_{ij} - \frac{k_ik_j}{2m}) \,\,.
\end{equation}

Maximizing modularity is an NP-complete problem \citep{brandes_modularity_2008}, making it computationally infeasible for large networks. As a result, a range of heuristics have been proposed, aiming to efficiently find highly modular partitions but without guaranteeing global optimality. These include greedy approaches such as the Clauset–Newman–Moore \citep{clauset_finding_2004} algorithm, and the multilevel Louvain \citep{blondel_fast_2008} and Leiden \citep{traag_louvain_2019} methods. In this work, we extend Louvain's multilevel modularity optimization technique with fairness constraints. 

\begin{figure}[t!]
    \centering
    \includegraphics[width=.95\linewidth]{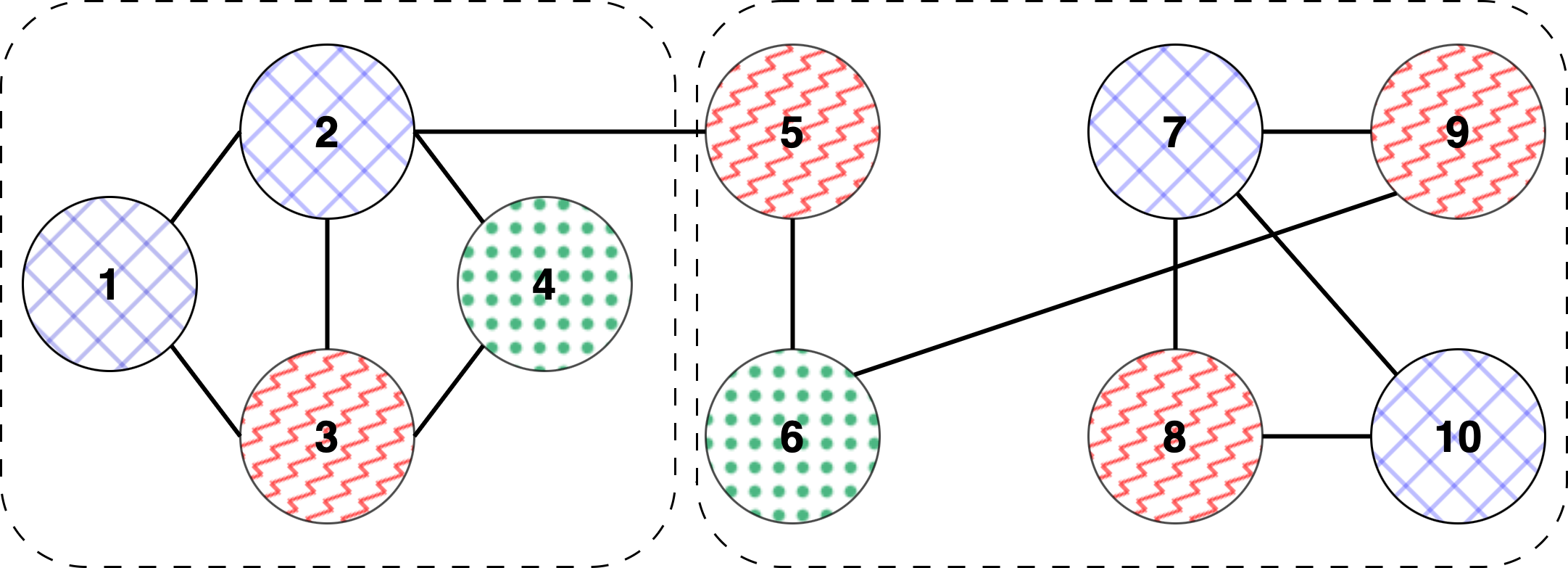}
    \caption{Example network with $V=\{1..10\}$. The set $H=\{\{1,2,7,10\},\{3,5,8,9\},\{4,6\}\}$ corresponds to the node membership in either of $K=3$ demographic groups coded as colors: blue (cross-hatch filling), red (diagonal filling) and green (dot filling)}
    \label{fig:fair-cd-example}
\end{figure}

\subsection{Problem statement}\label{ssec:problem}
We consider the following problem:
\problem{\textbf{(Weighted fairness-aware modular community detection)}}\label{prob:group-fair-cd}
Given an undirected weighted network $G=(V,E)$, a set $H$ denoting the membership of vertices in $V$ in $K \geq 2$ demographic groups (also referred to as colors), a fairness score $F_P$, and a weight $\alpha$, $0 \leq \alpha \leq 1$ controlling the trade-off between modularity and fairness, we want to obtain a partition $P$ of the vertices in $V$ into disjoint sets (called communities), maximizing a weighted sum of modularity and fairness. 
Namely, we want to optimize the following objective function:
\begin{equation}\label{eq:obj_func}
    \operatorname*{argmax}_P \, \, \, \alpha \cdot Q_{P} + (1-\alpha) \cdot F_{P} \,\,,
\end{equation}
where $Q_P$ is the modularity score for the partition $P$ (Eq. \ref{eq:modularity}) and $F_{P}$ is the global fairness score assigned to the partition $P$.

We illustrate Problem \ref{prob:group-fair-cd} above with an example. Consider the network $G$ in Fig. \ref{fig:fair-cd-example}, with ten nodes belonging to either of three color-coded demographic groups. Given a trade-off parameter $\alpha$, the goal is to partition of the vertices in $V$ into communities, e.g. $P=\{\{1..4\},\{7..10\}\}$ (in dashed lines).

This problem, however, comes with a fundamental empirical question: to what extent can both modularity and fairness simultaneously be optimized, and how can we obtain a good trade-off between the communities' quality and fairness? This depends on multiple factors: the choice of fairness score $F_P$, the value of $\alpha$, and the distribution of attribute values in $H$ with respect to the position of the edges in the input graph. We introduce the motivation behind our proposed proportionality-based fairness in the following section, and experimentally address the remaining aspects on both real and synthetic data in the remainder of this paper.

\section{Proportional balance fairness}\label{sect:fairness}
A common fairness definition in clustering is group balance, measuring how evenly different demographic groups (or colors) are represented within a community. Originally introduced by \citet{chierichetti_fair_2017}, it has been widely adopted in fairness-aware clustering applications, and later generalized to multiple groups \citep{bera_fair_2019,kleindessner_guarantees_2019,chhabra_overview_2021}.
Albeit natural and widely used, this definition is not well suited for fairness-aware community detection in networks with more than two groups. First, its scaling makes direct comparison across different numbers of groups difficult. Second, in networks with imbalanced group sizes, the maximum attainable balance score can be significantly below one, making it unclear whether low scores reflect unfairness or simply structural constraints.

To overcome these limitations, we introduce the proportional balance fairness score, a generalization of group balance that extends to any number of demographic groups, and normalizes for group size imbalances by introducing a penalty for communities where demographics are not proportionally represented. 
We begin by recalling the definition of group balance for a single community, as it forms the basis of our fairness function. We then introduce the expected proportional balance score for a community, before combining these into a single community fairness score, and, ultimately, a global fairness score for a partition.

\paragraph{Group balance}
First, we restate the multiple-group adaptation of group balance for communities as presented by \citet{panayiotou_towards_2025} for a single sensitive attribute:
\definition{\textbf{(Balance score)}}\label{def:comm-balance}
Given a community $C_i \subseteq V$ and a set $H$ denoting node membership in $K \geq 2$ demographic groups, the balance score for $C_i$ is given by:
\begin{equation} \label{eq:balance-score}
    balance(C_i) = (K-1) \cdot \min_{j \in [1..K]} \left(\dfrac{|H_j \cap V(C_i)|}{|H'_j \cap V(C_i)|}\right) \in [0,1] \,\,,
\end{equation}
where $V(C_i)$ are the vertices in community $C_i$, $H_j$ denotes the vertices in group $j$, and $H'_{j}$ the vertices not in group $j$.
We specifically consider this definition as it is scaled to receive a maximum balance score of one for perfectly balanced communities when $K>2$, since without scaling the maximum possible score becomes $1/(K-1)$.

\paragraph{Expected proportional balance}
However, simply considering the balance of colors in a single community can be problematic, particularly for heavily imbalanced networks where one demographic group is underrepresented. In extreme cases, the optimal balance score for a community can become much lower than the overall maximum of one, leading to potential issues in discovering balanced partitions. To address this limitation, we introduce a penalty for communities that do not follow the distribution of demographics in the network. To achieve that, we provide a score for the expected proportional balance of a community, if it follows the demographic distribution in the network. 

The rationale behind the expected proportional balance score is simple: if a community $C_i$ is large enough to contain at least one member from each demographic (i.e. $|V(C_i)|\geq K$), we calculate how many members of $C_i$ should be colored according to the ratio of demographics (which we also refer to as colors) in the entire network. Trivially, if a community is not large enough to represent all of the demographics, its balance score, and therefore, its expected proportional balance, becomes zero. 
After the first coloring step, there may remain some unaccounted members of $C_i$ that can belong to either group, with a probability equal to their ratio in the network. Should they belong in the least represented demographic, the balance score in $C_i$ is expected to increase, and vice versa for the remaining groups. 
As such, we define the expected proportional balance for a community as follows (see also Appendix \ref{app:prop-balance}): 

\definition{\textbf{(Expected proportional balance score)}}\label{def:prop-balance}
Given a community $C_i \subseteq V$ and a set $H$ denoting node membership in $K \geq 2$ demographic groups, the proportional balance score for $C_i$ is given by:
\begin{equation} \label{eq:proportional-balance}
    exp\_prop(C_i) =  
     \begin{cases} 
      0\,\,, & |V(C_i)|< K \\
      \dfrac{\phi \,K\,|V(C_i)| + (\phi+K-1-\phi K)\,n_e(C_i)}{[K\,|V(C_i)|+(\phi -1)\,n_e(C_i)]} \,\,, 
        & otherwise 
\end{cases}
\end{equation}
where $\phi$ is the balance score for the entire network, namely:
\begin{equation} \label{eq:phi}
    \phi = (K-1) \min_{j \in [1..K]} \left(\dfrac{|H_j \cap V|}{|H'_j \cap V|}\right) \in [0,1] \,\,, 
\end{equation}
and $n_e(C_i)$ is the count of members in $C_i$ not colored after the first step: 
\begin{equation} \label{eq:extra-nodes}
    n_e(C_i)=|V(C_i)| - \sum_{j \in [1..K]} 
    \left\lfloor\dfrac{|V(C_i)|\,|H_j \cap V|}{n}\right\rfloor \,\,.
\end{equation}

\paragraph{Proportional balance fairness score}
Next, we combine the previous scores into a proportional balance fairness score (simply referred to as proportional fairness hereinafter) for a single community, aiming to penalize communities that do not have a proportional representation of the demographic groups. A community $C_i$ receives a maximum score of one if its balance score is greater than or equal to the expected proportional balance score. Otherwise, we incur a penalty by subtracting the difference between the expected proportional balance and current balance scores for that community. In practice, a community that receives the maximum fairness score should be at least as balanced as a community following the distribution of groups in the network, if not improving the balance score.
We define the proportional fairness score for a single community as follows:

\definition{\textbf{(Proportional balance fairness score)}}\label{def:comm-fairness}
Given a community $C_i \subseteq V$, the proportional balance fairness score for $C_i$ is given by:
\begin{equation} \label{eq:comm-fairness}
    prop\_balance(C_i)=min(1,1-[exp\_prop(C_i)-balance(C_i)]) \in [0,1] \,\,.
\end{equation}

\begin{figure}[t!]
    \centering
    \includegraphics[width=.95\linewidth]{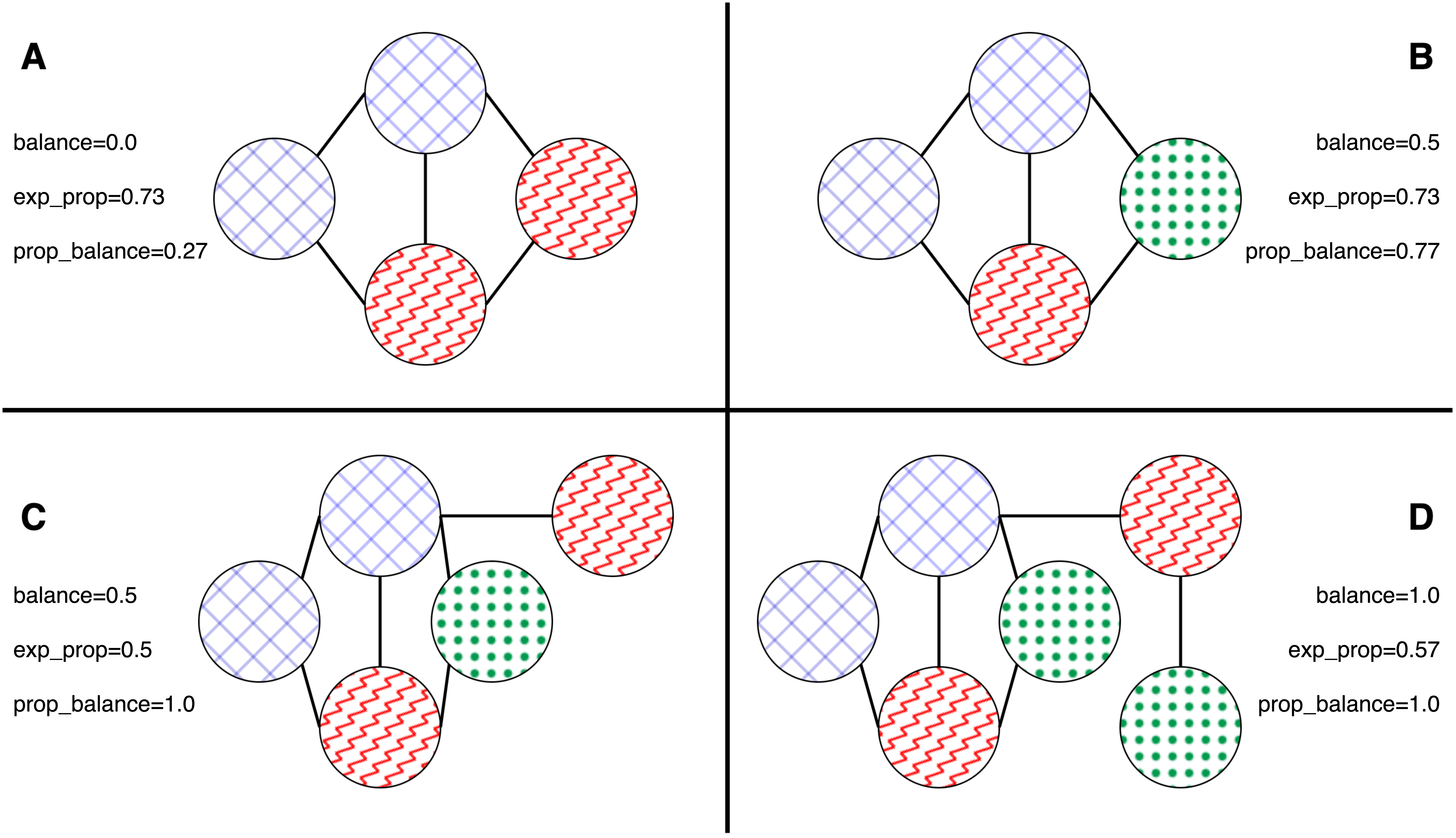}
    \caption{Examples of proportional balance fairness scores for various communities, rounded to two decimals. We consider the same network setting as Fig. \ref{fig:fair-cd-example} with $n=10$, $|H_1|=|H_2|=4, |H_3|=2, \phi=1/2$}
    \label{fig:prop-fairness-example}
\end{figure}

We illustrate the proportional fairness score with examples, considering the same network setting as the running example in Fig. \ref{fig:fair-cd-example}. Note that the $exp\_prop$ scores are only affected by the distributions of the sensitive groups and the size of the identified community. The proportional fairness score, on the other hand, penalizes communities where the balance score observed in the community is lower than the expected proportional balance for a community of that size; note that the final $prop\_balance$ score for the two communities depends on their observed balance score, with the unbalanced community of Fig. \ref{fig:prop-fairness-example}A receiving a far lower score than that in Fig. \ref{fig:prop-fairness-example}B. When the observed balance of that community follows the distribution of colors in the network (Fig. \ref{fig:prop-fairness-example}C), or exceeds that expectation (Fig. \ref{fig:prop-fairness-example}D), the community receives a maximum proportional fairness score of one.

\paragraph{Global fairness score}
Finally, following \citet{panayiotou_fair-mod_2025}, we define the global fairness score for the whole partition as the \textit{weighted average} of $prop\_balance(C_i)$ over all communities. Similar to our previous work, we argue that aggregating all individual communities' fairness scores into a single global score should consider not only the proportional balance of communities, but also their relative size. 
This prevents the formation of a large number of small, $K$-sized communities, in which each node has a different color. Although such a configuration would technically maximize the global fairness score under a simple average, it would not represent a meaningful or practically useful partition. Instead, by weighting communities according to their size, we ensure that fairness improvements in larger communities contribute more substantially to the global fairness score, reflecting both demographic proportionality and structural significance of the partition. 
Formally, the weighted global fairness score is defined as follows:
\definition{\textbf{(Global fairness score)}}\label{def:global-fairness}
Given an undirected, weighted network $G=(V,E)$ and a partition $P$ of $V$ into communities, the global fairness score for that partition is given by:
\begin{equation} \label{eq:global-fairness}
    F_{P} = \dfrac {
        \sum_{C_i \in P} \left(|V(C_i)| \, prop\_balance(C_i)\right)
    }{n} \in \left[0,1\right] \,\,.
\end{equation}

\begin{algorithm}
\caption{MOUFLON}
\label{alg:mouflon}

\begin{algorithmic}[1]

\Function {MOUFLON}{$G,\alpha,\theta$}
    \State $P \gets \{\{v\}: v \in V(G)\}$ \Comment{Assign each node into its own community}
    \State $P,opt' \gets$ \Call{Step1}{$G,P$} \Comment{Step 1: Local node movement on full graph}
    \Repeat
        \State $opt \gets opt'$
        \State $G \gets$ \Call{AggGraph}{$G,P$} \Comment{Partition becomes aggregated graph}
        \State $P \gets \{\{v\}: v \in V(G)\}$ 
        \State $P,opt' \gets $ \Call{Step2}{$G,P,\alpha$} 
        \item[] \Comment{Step 2: Local node movement on aggregated graph}
    \Until {$\theta \geq opt'-opt$} \Comment{Stop when gain $<\theta$}
    \State \Return {$P*$} \Comment{Yield flattened final partition}      
\EndFunction

\item[] 

\Function {Step1}{$G,P$}
    \Repeat
        \State $Q_{curr} \gets Q_P$ \Comment{Modularity for current partition (Eq. \ref{eq:modularity})}
        \For{$v \in V(G)$} \Comment{Visit nodes after random shuffle}
            \State $C' \gets \arg \max_{C \in P \cup \emptyset} \Delta Q_P(v \mapsto C)$ \Comment{Find best community for $v$}
            \If{$\Delta Q_{P}(v \mapsto C')>0$} \Comment{Only move for positive gains}
                \State $v \mapsto C'$ \Comment{Move node to best community}
            \EndIf
        \EndFor
    \Until {$Q_P \leq Q_{curr}$} \Comment{Stop when no more improvements}
    \State \Return{$P, Q_{curr}$}
\EndFunction

\item[] 

\Function {Step2}{$G,P,\alpha$}
    \Repeat
        \State $\mathcal{J}_{curr} \gets \mathcal{J}_{P,\alpha}$ \Comment{Score for current partition (Eq. \ref{eq:obj_func})}
        \For{$v \in V(G)$} \Comment{Visit nodes randomly}
            \State $C' \gets \arg \max_{C \in P \cup \emptyset} \Delta \mathcal{J}_{P,\alpha}(v \mapsto C)$ \Comment{Find best community}
            \If{$\Delta \mathcal{J}_{P,\alpha}(v \mapsto C')>0$} \Comment{Only move for positive gains}
                \State $v \mapsto C'$ \Comment{Move node to best community}
            \EndIf
        \EndFor
    \Until {$\mathcal{J}_{P,\alpha} \leq \mathcal{J}_{curr}$} \Comment{Stop when no more improvements}
    \State \Return{$P, \mathcal{J}_{curr}$}
\EndFunction

\item[] 

\Function {AggGraph}{$G,P$}
    \State $V' \gets P$ \Comment{Partitions become nodes in aggregate graph}
    \State $E' \gets \{(C_i,C_j) : (u,v) \in E(G), u \in C_i, v \in C_j, C_i, C_j \in P \}$ 
    \item[] \Comment{Connect communities in $P$ having edges between their nodes in $G$}
    \State $G' \gets$ \Call{Graph}{$V',E'$} 
    \State \Return{$G'$}
\EndFunction

\end{algorithmic}
\end{algorithm}

\section{MOUFLON: Multi-group Fairness-aware Louvain-based Community Detection}\label{sect:mouflon}
To tackle Problem \ref{prob:group-fair-cd} above, we propose MOUFLON (after Multi-grOUp Fairness-aware Louvain-based community detectiON), a scalable fairness-aware modular community detection method. This method aims to obtain a highly modular partitioning of the network, while also considering the previously defined group fairness constraints. 

MOUFLON is originally inspired by the greedy modularity optimization approach in the Louvain algorithm \citep{blondel_fast_2008}, which first assigns each node in the network to its own community, then iteratively merges neighboring communities if the move yields an overall positive gain in modularity. However, while simply replacing modularity gain with our objective function in Eq. \ref{eq:obj_func} can be effective for large values of $\alpha$, this method has limitations when prioritizing a more fair over modular partition. This effect is highlighted in \citet{panayiotou_fair-mod_2025}, where the Fair-mod method often cannot overcome local maxima for $\alpha < 0.5$. As the goal is to obtain a highly modular yet fair partition, we resolve this issue by first finding well-connected communities before adjusting the partition for fairness.

To address this problem, we propose a new heuristic, which we use as part of the MOUFLON method. To obtain a more modular partition, we move nodes locally as a first step, using only modularity as its objective function (cf. Alg.~\ref{alg:mouflon}, 12-23). Then, the algorithm performs subsequent moves on the aggregate graph based on the objective function in Eq. \ref{eq:obj_func}, considering both modularity and fairness gains (Alg.~\ref{alg:mouflon}, 24-35). We repeat the second step until the gain on the objective function does not surpass a threshold $\theta$, which we introduce to guard against performing additional iterations where merging communities yields extremely small gains.

Similar to Louvain, the time complexity of MOUFLON depends on the number of edges in the network and the number of iterations needed until convergence. However, calculating the potential fairness gains resulting from node movement can introduce a large overhead, especially during the first iterations when few nodes have been merged into smaller communities. To improve performance, we implement a hashtable-like data structure that tracks the demographic group proportions of each meta-node generated during the optimization process, to avoid repeated calculations of fairness score changes, introducing $\mathcal{O}(Kn)$ space complexity. With this improvement, the time complexity of each step remains roughly linear to the number of edges, with a fairness score calculation overhead depending on the number of demographic groups $K$ in the data, i.e. $\mathcal{O}(IKm)$, where $I$ is the number of iterations needed until convergence.

\section{Experimental evaluation}\label{sect:experiments}

In Table \ref{tab:fair-cd-comparison}, we identify desirable properties that a fairness-aware community detection method should satisfy. First, it should support multiple demographic groups and their distribution in the network, to avoid forming communities that are imbalanced towards any sensitive group in the data. Second, we need a mechanism to control the importance of community quality and the fairness score. Finally, the method should be non-parametric to the number of communities, and ideally scale to very large networks, to support the complexity and size of contemporary network data. 

By design, MOUFLON supports proportionality and multiple sensitive groups using our proposed community fairness definition, as well as scalability and non-parametricity to the number of communities, as it uses a modularity-based optimization technique inspired by the Louvain algorithm. 
However, the latter properties, namely quality-fairness trade-off, number of communities and scalability, are largely affected by the size, density and structure of the network, but also the distribution of sensitive groups, since it affects the fairness scores that can be achieved for that network. Therefore, these aspects of the algorithm should be investigated experimentally.

In the following, we evaluate how the aforementioned properties affect the behavior of our method, under both real and synthetic data. Specifically, we focus on the following questions:
\begin{itemize}
    \item \textbf{Q1:} How is performance affected by network size and density?
    \item \textbf{Q2:} How are the identified communities affected by $\alpha$?
    \item \textbf{Q3:} How are the identified communities affected by the distribution of the sensitive groups?
    \item \textbf{Q4:} How are the identified communities affected by the fairness metric?
    \item \textbf{Q5:} How are the identified communities affected by highly clustered homogeneous groups?
\end{itemize}

\subsection{Datasets}\label{ssec:datasets}
\paragraph{Synthetic data}
We use a variety of synthetic datasets to evaluate our method. To assess the effect of the minority demographic group on our method, we consider synthetic networks with two demographic groups (which we refer to as blue and red nodes).

First, to assess the method's scalability, we generate random Erd\H{o}s-R\'enyi (ER) networks of varying sizes, with respect to the number of both nodes and edges. We color the nodes independently, with equal probability of belonging to either color.
Specifically, we consider the following settings, where $p$ is the probability of edge generation:
\begin{itemize}
    \item (1a) ER network, $n=[1000,\ldots,200000]$, $p=0.001$
    \item (1b) ER network, $n=10000$, $p=[0.1,\ldots,0.5]$
\end{itemize}

Second, we want to study the differences in partition quality and fairness outcomes with respect to different network structures. 
In this case, instead of simply generating random Erd\H{o}s-R\'enyi networks, we use a clique-based generator with a probability of edge rewiring. This is to have a reasonable expectation of the number of communities identified by the algorithm.

The network generator produces $L$ cliques of size $l$, and then rewires edges so that they connect different cliques, with a probability $p\_rewire$. All nodes in the network are then colored with a probability given in vector $p\_groups$. To further evaluate the partition quality, we use two strategies to color nodes: (1) color nodes individually according to $p\_groups$, (2) color entire cliques according to $p\_groups$. 
For consistency between experiments, the network structure remains the same throughout all generated synthetic networks.
As we are generating synthetic nodes under two demographic groups, we denote the node coloring probability for the minority group (i.e. the smallest probability in all elements of $p\_groups$) as $p\_sensitive$. 

We generate rewired-clique networks under the following settings:
\begin{itemize}
    \item (2a) Rewired clique network with individual coloring, $L=100$, $l=10$, $p\_rewire=0.1$, $p\_sensitive=[0.1,\ldots,0.5]$,
    \item (2b) Rewired clique network with clique coloring, $L=100$, $l=10$, $p\_rewire=0.1$, $p\_sensitive=[0.1,\ldots,0.5]$.
\end{itemize}

\paragraph{Real-world data}
We also evaluate MOUFLON on real-world social networks varying in size, sourced from the Stanford Large Network Dataset Collection \footnote{\url{https://snap.stanford.edu/data/}}. Specifically, we select the following networks:
\begin{itemize}
\item \textit{Facebook}: The dataset includes Facebook friend lists of survey participants. We consider the users' listed gender as the sensitive attribute (anonymized in the dataset).
\item \textit{Deezer}: The social network consists of European users of the Deezer online social network, where an edge between two users means they both follow the same artist online. The investigated sensitive attribute demographic is the users' gender (anonymized).
\item \textit{Twitch Gamers}: The network (referred to simply as Twitch hereinafter) includes users on the Twitch streaming platform, and their mutual streamer following relationships. We color the nodes according to whether their stream is listed as appropriate for mature audiences.
\item \textit{Pokec}: The dataset consists of users of the Pokec online social network, and friendships on the site between them. For this dataset, we consider both age (\textit{Pokec-a}) and gender (\textit{Pokec-g}) of the users as sensitive attributes. Similarly to \citet{panayiotou_fair-mod_2025}, for the age attribute, we split the nodes into red and blue according to the median age, after removing nodes without an age value set in their profile. Gender information is anonymized in the dataset.
\end{itemize}
A summary of the network characteristics can be seen in Table \ref{tab:real-sn}.

\begin{table}[t!]
\centering
\resizebox{.6\linewidth}{!}{
\begin{tabular}{lrrr}
\toprule
\textbf{Dataset} & \textbf{$n$} & \textbf{$m$} & \textbf{$\phi$} \\
\midrule
Facebook & 4,039 & 88,234 & 0.614 \\
Deezer & 28,281 & 92,752 & 0.796 \\
Twitch & 168,114 & 6,797,557 & 0.887 \\
Pokec-a & 1,138,314 & 10,794,057 & 0.825 \\
Pokec-g & 1,632,640 & 22,301,602 & 0.971 \\
\bottomrule
\end{tabular}
}
\caption{Real-world network characteristics: $n$ and $m$ are the network size in nodes and edges, respectively, and $\phi$ is the overall balance score in the network (Eq. \ref{eq:phi})}
\label{tab:real-sn}
\end{table}

\subsection{Settings}\label{ssec:settings}
We implement the MOUFLON method as described in Algorithm \ref{alg:mouflon} using Python, and utilizing graph objects from the NetworkX library. 
We conduct the experiments on a typical desktop environment; specifically, we use a virtual machine running Ubuntu 22.04, with 8 cores at 2.1 GHz and 32GB RAM.

To address uncertainty, including the random shuffle behind our algorithm, we report the average value and standard deviation for the modularity and fairness scores, number of communities identified, and execution time, over multiple runs. We repeat all the synthetic data experiments ten times, and the real-world network experiments three times.

\subsection{Results}\label{ssec:results}
In this section, we present our evaluation of MOUFLON. Our experiments are guided by Questions 1-5 outlined above. First, we evaluate the method's scalability in terms of execution time. Second, we study how the identified communities are affected by different algorithmic choices: the hyperparameter $\alpha$, which controls the quality-fairness trade-off, and the measure for quantifying partition fairness. Finally, we examine how community quality and fairness are affected by structural properties of the network, such as the distribution of sensitive groups and the presence of homogeneous, tightly connected groups.

\begin{figure}[t!]
\centering
\includegraphics[width=\textwidth]{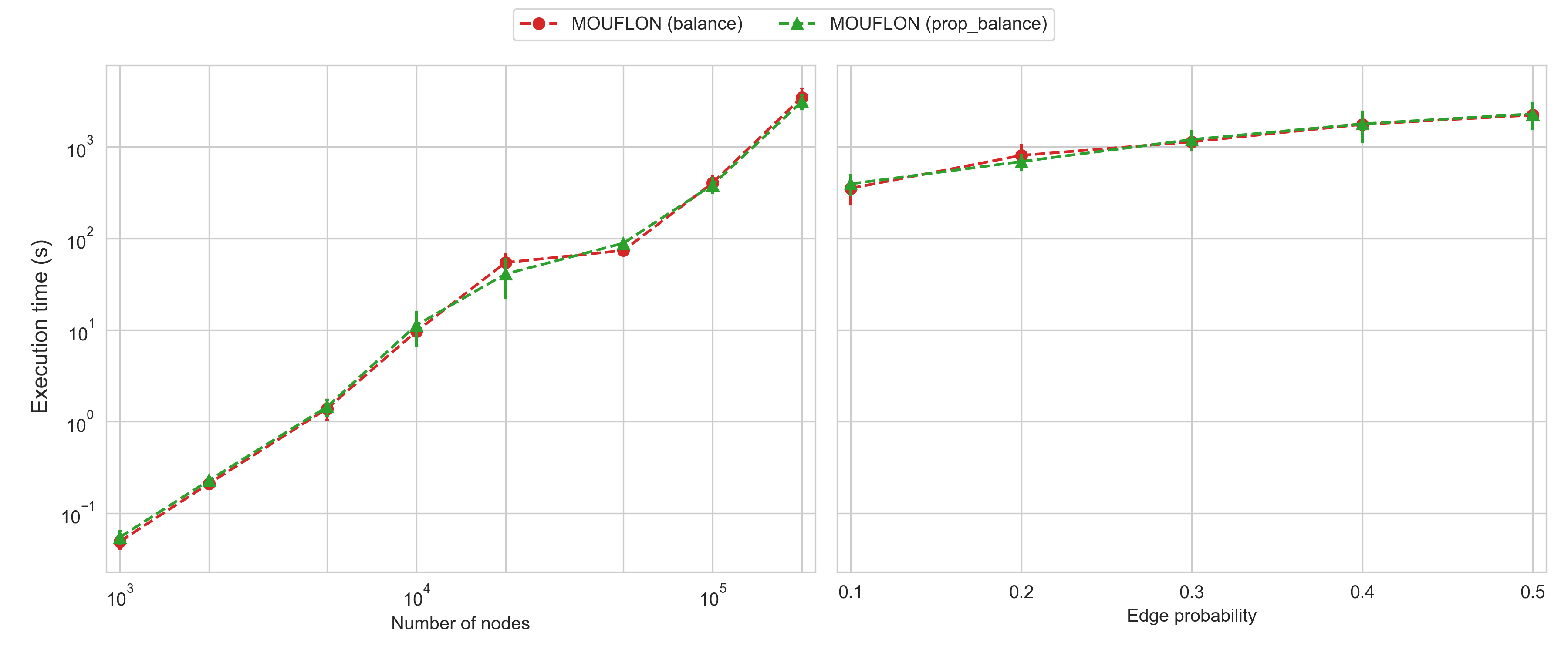}
\caption{Execution time of MOUFLON, $\alpha=0.5$ (in seconds). 
Left panel: synthetic data (1a) with increasing network size $n$, $p=0.001$.
Right panel: synthetic data (1b) with increasing network density, $n=10000$}
\label{fig:scalability_synth}
\end{figure}

\begin{table}[t!]
\centering
\resizebox{\textwidth}{!}{
\begin{tabular}{
l
S[table-format=2.1(1.1)]
S[table-format=3.1(1.1)]
S[table-format=4.1(3.1)]
S[table-format=4.1(3.1)]
S[table-format=5.1(4.1)]
}
\toprule
\textbf{Method (Fairness metric)} & \textbf{Facebook} & \textbf{Deezer} & \textbf{Twitch} & \textbf{Pokec-a} & \textbf{Pokec-g} \\
\midrule
MOUFLON (balance) & 0.9(0.1) & 4.6(0.3) & 214.7(49.6) & 1195.3(237.8) & 2193.7(348.4) \\
MOUFLON (prop\_balance) & 1.0(0.1) & 4.6(0.6) & 194.7(26.7) & 1128.9(290.1) & 2462.3(417.6) \\
\bottomrule
\end{tabular}
}
\caption{Execution time of MOUFLON for real-world networks. We report the average execution time in seconds and standard deviation over three runs, $\alpha=0.5$}
\label{tab:scalability-real}
\end{table}

\paragraph{Q1: How is performance affected by network size and density?}
To evaluate scalability, we measure execution time as network size and density increase, using synthetic ER-network datasets and the real-world social networks listed in Table \ref{tab:real-sn}. 

As shown in Fig. \ref{fig:scalability_synth}, MOUFLON follows a roughly log-linear growth pattern in execution time with increasing network size. We observe a relatively large standard deviations for $n=10000$ and $n=20000$, which can be attributed to the random node shuffling during the initialization of each step, which affects the order of merging the nodes when forming communities.
Similar growth trends are observed when increasing network density. As expected, replacing the fairness metric from simple group balance to proportional balance results in minor differences in execution time.

The performance of these methods on real datasets (Table \ref{tab:scalability-real}) confirms the general trends observed in synthetic data. We again observe that increasing network density has a clear impact on execution time; note the large performance differences between the age and gender variants of the Pokec dataset. However, MOUFLON also shows relatively large standard deviations, particularly for the larger Pokec networks, a result of the random node shuffling at the beginning of each step of the optimization process.

\begin{figure}[t!]
\centering
\includegraphics[width=\textwidth]{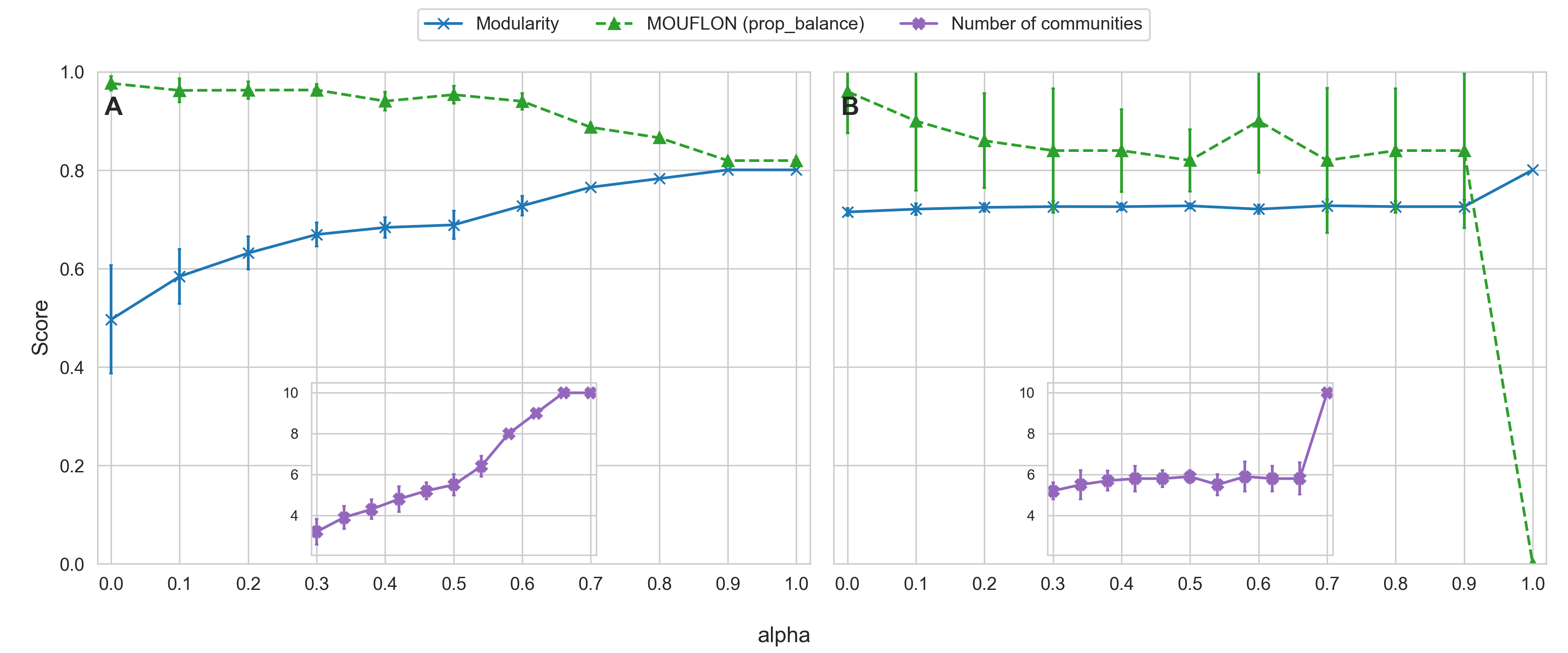}
\caption{Modularity and fairness scores for the default MOUFLON (prop\_balance) method on synthetic data with $p\_sensitive=0.5$, for varying $\alpha$, and number of communities (inset panels) of the identified partition. (A) Synthetic data with individual node coloring (2a). (B) Synthetic data with clique coloring (2b)} 
\label{fig:quality_mouflon_alpha}
\end{figure}

\paragraph{Q2: How are the identified communities affected by $\alpha$?}
We next analyze how the trade-off hyperparameter $\alpha$ affects modularity, fairness, and the number of communities identified by MOUFLON. 
Fig. \ref{fig:quality_mouflon_alpha}A shows the results for synthetic clique-based networks, where both demographic groups are equal in size and randomly distributed. 

First, we note that increasing $\alpha$ leads to higher community quality scores (modularity) and lower fairness (proportional balance). For $\alpha=1$, MOUFLON behaves similarly to Louvain, as we only optimize for modularity. This outcome matches our initial expectations, since larger values of $\alpha$ are designed to favor more modular partitions.
Because the two demographic classes are equally distributed in the network when $p\_sensitive=0.5$, the balance and proportional balance scores for the partition are the same.
When prioritizing community fairness over quality, particularly when $\alpha \leq 0.3$, we observe relatively large variations in the modularity score obtained. This is a result of the presence of multiple possible partitions that can receive a similar high fairness score; note the small standard deviation on the fairness scores for runs with $\alpha \leq 0.3$. 

Second, we observe that the number of the identified communities increases with $\alpha$, eventually reaching ten for $\alpha \geq 0.9$. This finding also aligns with our initial expectations, as the most modular partition, identified by a Louvain-like method, corresponds to the ten initially planted cliques in the synthetic network data as their own communities.
Experiments on the real social networks broadly confirm the previously observed trends: modularity scores and the number of identified communities increase with $\alpha$, while the fairness scores decrease (cf. also Appendix \ref{app:real-data}). 

\begin{figure}[t!]
\centering
\includegraphics[width=\textwidth]{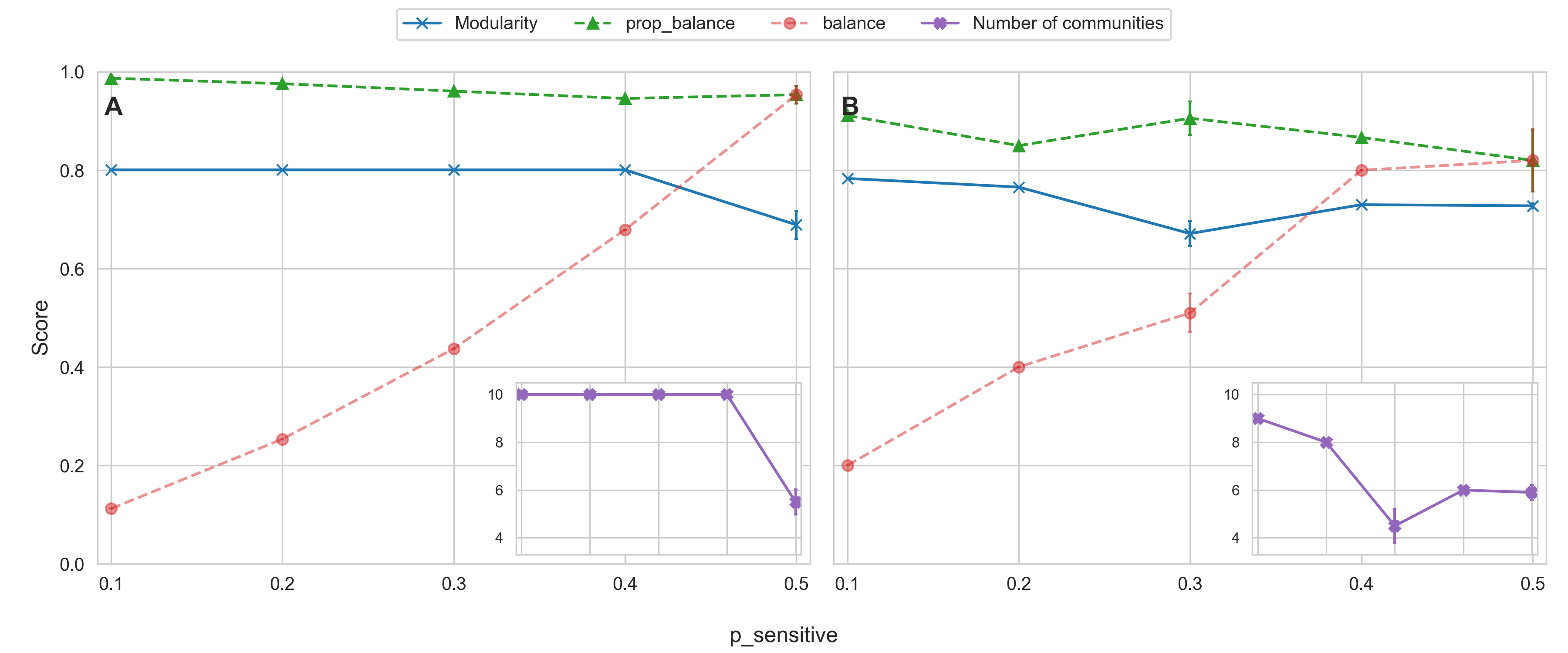}
\caption{Modularity and fairness scores for the default MOUFLON (prop\_balance) method on synthetic data, $\alpha=0.5$, for varying $p\_sensitive$, and the number of communities (inset panels) for the identified partition. (A) Synthetic data with individual node coloring (2a). (B) Synthetic data with clique coloring (2b)} 
\label{fig:quality_mouflon_psensitive}
\end{figure}

\paragraph{Q3: How are the identified communities affected by the distribution of the sensitive groups?}
In Fig. \ref{fig:quality_mouflon_psensitive}A, we examine the effect of group imbalances on MOUFLON by varying the proportion of nodes belonging to the sensitive group ($p\_sensitive$). 
As expected, the proportional balance score remains high (and close to one) regardless of the demographic distribution, with minor fluctuations attributed to the randomness in node coloring during the generation of the synthetic networks. In stark contrast, the balance score for the same partitions increases approximately linearly with $p\_sensitive$. This is because the demographic distribution of the identified communities is roughly proportional to the overall distribution in the network.

For $p\_sensitive \leq 0.4$, the identified partitions contain ten communities corresponding to the initially generated cliques; note that the modularity score remains at $0.8$. Interestingly, however, for $p\_sensitive=0.5$ we observe a drop in the modularity score, a result of the algorithm merging parts of the generated cliques to identify a partition that is more fair than modular (as $\alpha = 0.5$).

\begin{figure}[t!]
\centering
\includegraphics[width=\textwidth]{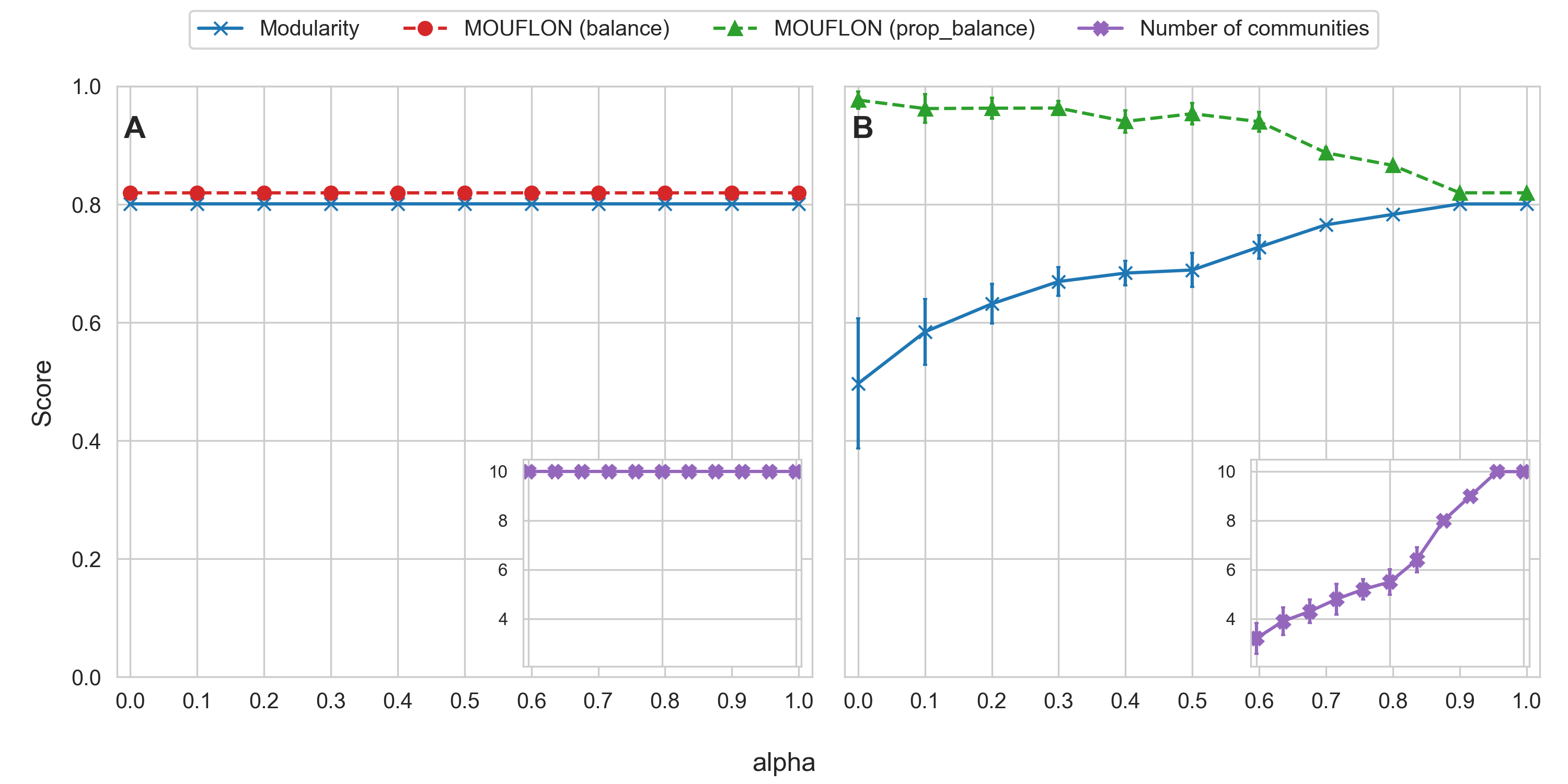}
\caption{Modularity and fairness scores for MOUFLON using (A) balance, (B) prop\_balance as the fairness metric. Experiments are performed on synthetic data with individual node coloring (2a) and $p\_sensitive=0.5$, for varying $\alpha$. Inset panels show the number of communities} 
\label{fig:quality_mouflon_plus_alpha}
\end{figure}

\begin{figure}[t!]
\centering
\includegraphics[width=\textwidth]{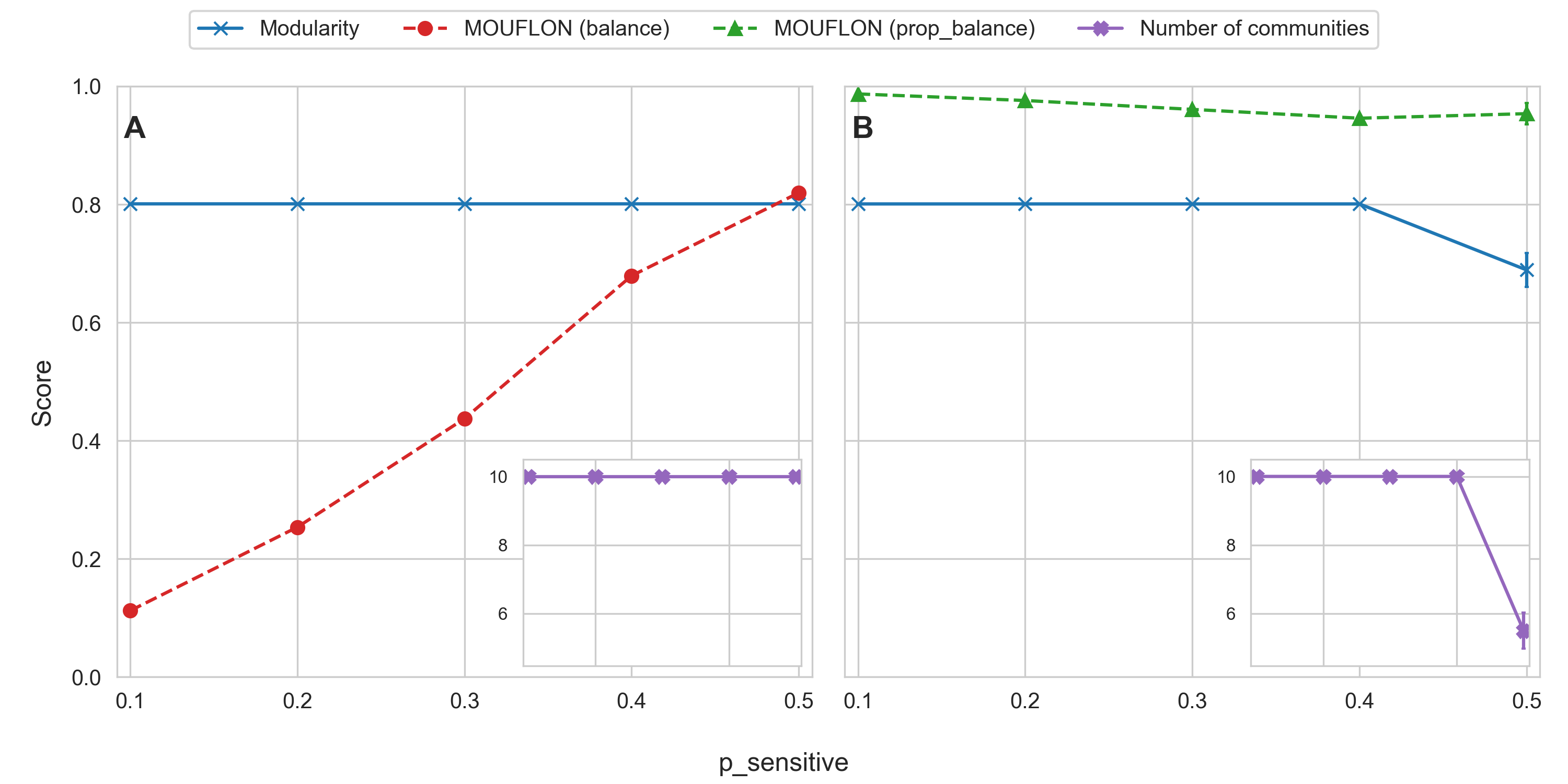}
\caption{Modularity and fairness scores for MOUFLON using (A) balance, (B) prop\_balance as the fairness metric. Experiments are performed on synthetic data with individual node coloring (2a) and $\alpha=0.5$, for varying $p\_sensitive$. Inset panels show the number of communities} 
\label{fig:quality_mouflon_plus_psensitive}
\end{figure}

\paragraph{Q4: How are the identified communities affected by the fairness metric?}
Next, we compare the differences in the identified communities when using group balance as the fairness metric in MOUFLON; that is, calculating the balance for each community in the partition under Def. \ref{def:comm-balance}, and aggregating to a global fairness score using the weighted average method of Eq. \ref{eq:global-fairness}.

Plotting the average fairness and modularity scores for the partitions obtained in Fig. \ref{fig:quality_mouflon_plus_alpha} leads to another interesting observation: using simple group balance as the fairness metric in MOUFLON leads to partitions that are not sensitive to changes in $\alpha$. Notably, the modularity and fairness scores remain exactly the same regardless of $\alpha$, as the first optimization step identifies the ten planted cliques as communities, and the algorithm is later unable to overcome that local maximum. In turn, this shows that proportional fairness is a more appropriate fairness metric for this problem, as it allows adjusting the importance of network structure over fairness. 

In Fig. \ref{fig:quality_mouflon_plus_psensitive} we note that adjusting $p\_sensitive$ leads to limited differences in the partitions identified. In most cases, with only one exception also seen in Fig. \ref{fig:quality_mouflon_psensitive}, the ten cliques are assigned as their own communities. The only difference is seen at the linear increase of the balance fairness score, while proportional balance remains roughly similar regardless of $p\_sensitive$. 

The aforementioned effects also remain similar when testing the method on real networks (cf. Appendix \ref{app:real-data}). There, we note the ability for MOUFLON to better overcome the local maxima issue pointed out in \citep{panayiotou_fair-mod_2025}, while we also see how our method using the default proportional balance as the fairness definition is more sensitive to changes in $\alpha$. 
Similarly, we note the tendency of our method using proportional balance to create fewer, larger communities for smaller values of $\alpha$, before reducing them to multiple smaller communities as $\alpha$ increases. The opposite effect is visible when replacing the fairness metric from proportional balance to simple balance.

\paragraph{Q5: How are the identified communities affected by highly clustered homogeneous groups?}
Finally, we investigate how strongly clustered groups affect the communities that can be identified by our method. In Fig. \ref{fig:quality_mouflon_alpha}B, we consider the extreme scenario where entire cliques are monochromatic, that is, all of the nodes belong to a single demographic group. Notably, the modularity score remains roughly around 0.75 for any $\alpha<1$, whereas the proportional balance fairness score is largely affected by the random node shuffle and the order in which meta-nodes are merged to form communities, as indicated by the large variations in the fairness scores. Fairness scores then decrease to zero when the cliques are assigned as their own communities for $\alpha=1$.

For the monochromatic clique-based synthetic networks in Fig. \ref{fig:quality_mouflon_psensitive}B, we observe several interesting behaviors of our method. First, we note the tendency to merge two monochromatic cliques together as single communities to maximize the fairness score, as $p\_sensitive$ increases. The only exception can be seen for $p\_sensitive=0.3$, where the edge rewiring procedure behind the synthetic network generation forces multiple cliques merged into a single one to maximize fairness.
Second, despite the linear increase trend for the balance score remaining broadly similar to the individual coloring dataset (cf. Fig. \ref{fig:quality_mouflon_psensitive}A), we observe fluctuations around $0.9$ for the proportional balance score.

\section{Discussion}\label{sect:discussion}
The experiments above lead to various interesting insights on implementing fairness constraints in modularity-based community detection.
First, we note that MOUFLON, with its modularity-first heuristic, is able to overcome the local maxima issues identified by \citet{panayiotou_fair-mod_2025}. 
We observe that generally, in both the synthetic and real-world networks, increasing $\alpha$ leads to a more modular partition and more communities. In contrast, lower values of $\alpha$ prioritize partitions with higher fairness scores, that are still highly modular as a result of the modularity-first optimization approach of our method.

Second, we observe how the fairness function chosen largely influences both the quality of communities that can be identified, and the trade-off between modularity and fairness outcomes. 
Varying the proportion of sensitive group membership ($p\_sensitive$) shows how the newly proposed proportional balance score remains stable regardless of class imbalances, in contrast to simple group balance within a community. 
Among the two pattern-based metrics examined, only proportional balance enables tunable trade-offs between modularity and fairness via $\alpha$. MOUFLON under simple balance fairness remains insensitive to the $\alpha$ hyperparameter, suggesting limited flexibility when optimizing for different trade-offs of partition quality versus fairness outcomes using our method.

Matching our initial expectations, using the proposed proportional balance fairness metric, we can discover partitions with high fairness scores, as any class imbalances in the network data have very small effects on the fairness scores, and in turn, the identified communities. Indeed, one of the main aims behind this fairness metric is to obtain a normalized score which is sensitive to demographic imbalances, as simple balance measures are not appropriate when a group is overrepresented in the data \citep{panayiotou_towards_2025}.
This comes with another interesting observation: when prioritizing partition fairness, MOUFLON tends to merge together strongly connected groups that would otherwise be identified as their own communities, often yielding fewer communities in the partition than when increasingly prioritizing modularity. This finding is in contrast with the effects noted in \citet{panayiotou_fair-mod_2025}, where partitions prioritizing fairness under smaller values of $\alpha$ contain more and smaller communities.

Furthermore, we also note that how the sensitive groups are connected to each other can affect the types of partitions identified: our monochromatic clique-based synthetic data reveal that strongly segregated group structures create rigid partitions, where improving their fairness would require significantly compromising modularity. 
In the synthetic network set where nodes have been randomly colored, we observe a smooth optimization between fairness and partition quality; we note the expected trade-off between network structure and demographic balance as $\alpha$ increases, alongside the convergence of the number of communities to the number of planted cliques. In contrast, when groups are highly clustered, fairness becomes much harder to optimize; modularity remains stable while fairness scores fluctuate substantially. 
This result emphasizes the need for novel definitions of fairness that consider multiple demographic groups and the edges between them, along with extensive benchmarking of fairness metrics in community detection, and whether they can be maximized under different optimization objectives and network structures. 

In terms of scalability, we note that the current implementation of MOUFLON takes seconds to partition small-to-medium-size networks (in order of $10,000$ nodes and $50,000$ edges), scaling to a few minutes for larger networks, and is also able to yield communities in very large networks in tens of minutes. While the efficiency of the current Python implementation is limited by the usage of graph objects in the Python-native NetworkX library to represent the underlying network data, these results show that MOUFLON can already be used in practice to cluster very large networks.

When interpreting the aforementioned results, it is important to consider several aspects also pertaining to the Louvain community detection method. First, since the initialization requires shuffling the nodes, the obtained partition might differ between executions. In our evaluation, we address this limitation by providing average results and standard deviation over multiple executions.
Second, the modularity score strongly depends on the network structure, and can thus vary greatly depending on network sparsity. This highlights the importance of experimentally setting the hyperparameter $\alpha$ to an appropriate value. Moreover, this implies that the objective score for the partition is not comparable across different networks.

Additionally, as modularity maximization is an NP-complete problem \citep{brandes_modularity_2008}, 
our heuristic, adapting the Louvain approach designed to address Problem \ref{prob:group-fair-cd}, both provides an effective solution, and mitigates the local maxima issues observed in \citet{panayiotou_fair-mod_2025}. However, while this modularity-first heuristic can identify partitions with high modularity before then adjusting meta-nodes for fairness, it constrains the decomposition of meta-nodes, in specific cases where moving a subset of nodes to other communities would improve fairness. Although this limitation lies beyond the scope of the present study, future work should explore and benchmark alternative modularity-based heuristics considering this issue.

Finally, an important note concerns the fairness definition itself, as the choice behind what should be considered a fair partition varies depending on the use case and network type. In this work, we consider a definition based on the balance between nodes in a community, which is a well-studied definition of fairness that is extensible to multiple demographics. 
Other recently proposed fairness metrics, such as the modularity-based metrics of \citet{manolis_modularity-based_2023} and \citet{gkartzios_fair_2025}, capture fairness based on the connections between individuals, which might be more suitable when considering more complex network types, such as multilayer networks. However, integrating multiple sensitive attributes and considering intersectional subpopulations remain open questions within this context. Considering the type of community detection method can also affect fairness via the number and size of communities identified \citep{de_vink_group_2025,de_vink_quantifying_2025,panayiotou_towards_2025}, it is critical to consider how the fairness definition and community quality in synergy can affect fairness outcomes.

Overall, our experimental evaluation highlights the need to carefully consider the effects of several aspects when evaluating fairness-aware network analysis methods: the fairness definition and the proportion of demographic groups present in the network, but also the network structure itself and how these demographics are grouped together. This can be done by designing synthetic networks mindful of how these properties can affect the fairness outcomes of the algorithm.  

\section{Conclusion}\label{sect:conclusion}
Integrating fairness constraints into community detection is crucial for systems like online social media, where traditional clustering methods risk reinforcing echo chambers and latent biases encoded in the data.
In this work, we propose MOUFLON, a scalable, multi-group fairness-aware community detection method with a novel fairness metric that better reflects demographic balance across various network structures, including networks where demographic groups are imbalanced. 
We evaluate the method's performance and the trade-off between modularity and fairness, focusing on various properties of the data and algorithm: network size, network density and fairness metric. 
Our experimental evaluation also considers various often overlooked aspects, such as the impact of the sensitive group distribution and highly clustered homogeneous groups in fairness outcomes. 
Our results provide useful insights into community detection with fairness constraints, both showcasing the impact of the aforementioned network properties, and underscoring the importance of thoughtful benchmark design when evaluating fairness-aware network analysis methods in general.

While in this work we focus on online social networks, community detection incorporating fairness constraints can prove useful in other domains. Such examples include randomized experiments requiring balanced communities \citep{saveski_detecting_2017}, and educational examples of assigning school groups while simultaneously considering the demographics such as gender and ethnicity \citep{kroneberg_classroom_2021,kruse_re-print_2024}.
Moreover, transportation-related applications can also benefit from such fairness-aware methods, as community detection is often used on both social and spatial network data in the design of mobility networks \citep{de_montis_commuter_2013,yan_transit_2023,guo_f-deepwalk_2024}.

Future work on fairness-aware community detection should consider both alternative modularity-based optimization methods other than the Louvain multi-level approach, for example initially assigning sensitive nodes in communities before optimization begins \citep{viles_constrained_2023}, and integrating fairness constraints into optimization functions other than modularity, as fairness outcomes can be affected by the type of method chosen \citep{de_vink_group_2025,de_vink_quantifying_2025,panayiotou_towards_2025}. 
Moreover, alternative definitions of fairness should be proposed, in order to promote fairness for intersectional subgroups and multiple demographics simultaneously \citep{martin-gutierez_intersectional_2024,panayiotou_towards_2025}, as network data is often enriched with multiple demographic attributes. 

Finally, considering the growing availability of feature-rich, large-scale social network data, fairness-aware methods should be extended to more complex network models, such as multilayer networks \citep{van_der_laan_whole_2023,bokanyi_anatomy_2023,kazmina_socio-economic_2024,cremers_unveiling_2025,panayiotou_anatomy_2025}. This direction is particularly important, both because of their popularity in modeling large social networks with multiple types of ties and demographic attributes, and due to the additional computational challenges when multiple layers are considered simultaneously \citep{panayiotou_current_2024}.

\section*{Statements and Declarations}

\subsection*{Funding}
GP has been partly funded by eSSENCE, an e-Science collaboration funded as a strategic research area of Sweden.

\subsection*{Competing interests}
The authors declare no competing interests.

\subsection*{Author contributions}
GP has designed the algorithm and fairness metric, and led writing of the paper. AMMS and GP have developed the code, performed the computational data analysis, and prepared the visualizations. GP, MM and EC have supervised the research. All authors have formulated research goals and aims, and have edited and reviewed the manuscript.

\subsection*{Data availability}
The real-world datasets analyzed are publicly available through the Stanford Large Network Dataset Collection. The implementation of MOUFLON, along with the synthetic network generator, are available at \url{https://github.com/uuinfolab/paper.25_DKMD_MOUFLON}.

\subsection*{Acknowledgments}
The computations were enabled by resources provided by the National Academic Infrastructure for Supercomputing in Sweden (NAISS), partially funded by the Swedish Research Council through grant agreement no. 2022-06725.
We would like to thank Georgios Fakas for useful remarks on early versions of the manuscript, and Xin Shen for providing a first version of the clique-based synthetic network generator.

\begin{appendices}
\counterwithout{figure}{section}
\setcounter{figure}{7}

\section{Expected proportional balance score}\label{app:prop-balance}
In this appendix, we provide more details behind the calculation of the expected proportional balance score. The metric represents an expectation of the balance score of a community whose members' demographics roughly follow the distribution of groups in $H$, and thus, the overall balance score $\phi$ if the entire network is assigned as a single community (Eq. \ref{eq:phi}).
We specifically focus on the second case, where the size of a community $C_i$ is larger than the number of demographics in the network, i.e. $|V(C_i)| \geq K$.

To calculate this score, we first color enough nodes in $C_i$ to roughly follow the ratio of color membership in $H$. The nodes in proportion are given by
\begin{equation*}\label{eq:prop-nodes}
n_p(C_i) = \sum_{j \in [1..K]} 
    \left\lfloor\dfrac{|V(C_i)|\,|H_j \cap V|}{n}\right\rfloor .
\end{equation*}

Out of the remaining $n_e(C_i)$ non-colored nodes, where $n_e(C_i)=|V(C_i)|-n_p(C_i)$, we assume that each node belongs to either group with an identical probability $1/K$.
If the extra node belongs to the least represented demographic in $G$, the balance score for that community (Eq. \ref{eq:balance-score}) increases, as the numerator is based on the group with the fewest members, while the denominator remains the same, assuming all groups in $H$ do not contain the same amount of members. 
Conversely, if they belong to any other group (with a probability $(K-1)/K$), the balance score would decrease.

As a result, the expected proportional balance score for a proportionally colored community $C_i$ (when $|V(C_i)| \geq K$) becomes:
\begin{equation*}\label{eq:full-prop-balance}
\begin{split}
    exp\_prop(C_i)
    & = \dfrac{(K-1)\left[\dfrac{\phi\, n_p(C_i)}{\phi+K-1}+\dfrac{n_e(C_i)}{K}\right]}{\dfrac{(K-1)\,n_p(C_i)}{\phi+K-1}+\dfrac{(K-1)\,n_e(C_i)}{K}} \\
    & = \dfrac{(K-1) \left[\phi \,K\,[|V(C_i)|-n_e(C_i)] + (\phi+K-1)\,n_e(C_i) \right]}{K(K-1)[|V(C_i)|-n_e(C_i)]+(\phi+K-1)(K-1)\,n_e(C_i)} \\
    & = \dfrac{\phi \,K\,|V(C_i)| + (\phi+K-1-\phi K)\,n_e(C_i)}{K\,|V(C_i)|+(\phi -1)\,n_e(C_i)}
\end{split}
\end{equation*}

\begin{figure}[t!]
    \centering
    \includegraphics[width=.75\linewidth]{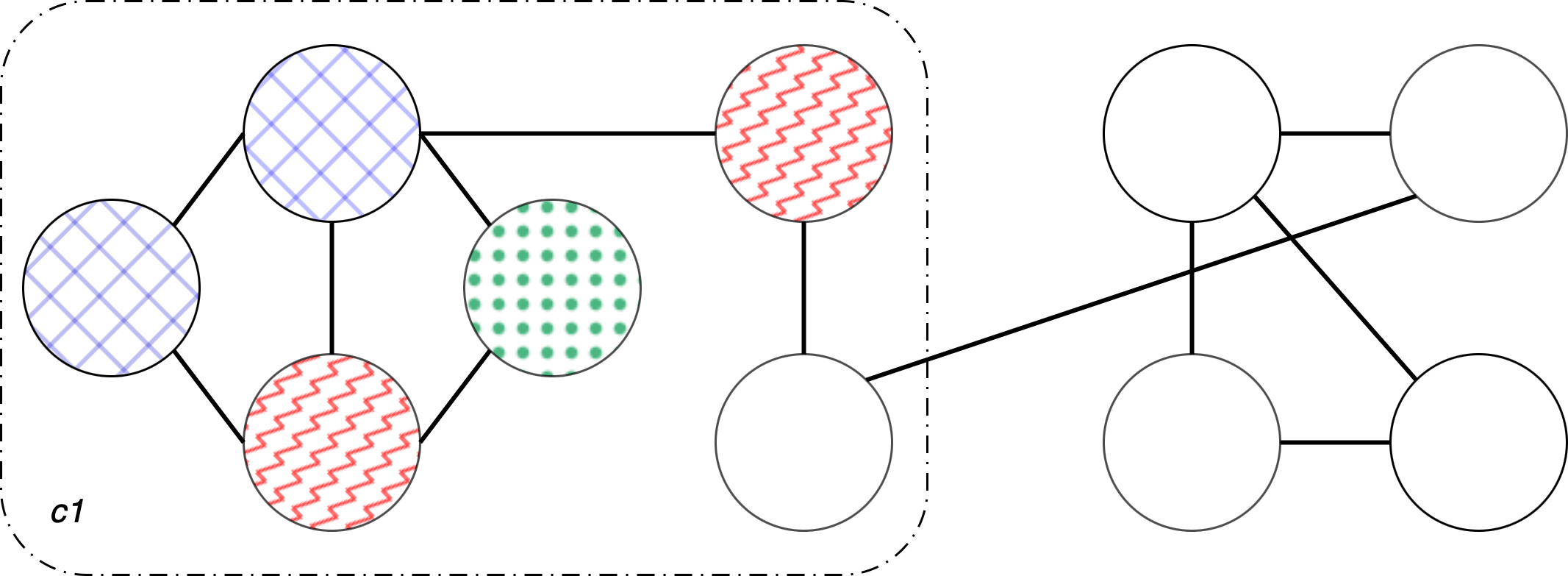}
    \caption{Example network from Fig. \ref{fig:fair-cd-example}, with a community $c_1$ identified (dashed line), after proportionally coloring its nodes}
    \label{fig:prop-balance-example}
\end{figure}

We further illustrate this calculation with an example. Consider again the example network from Fig. \ref{fig:fair-cd-example} with $|V|=10$, $|H_1|=4$ blue (cross-hatch filling), $|H_2|=4$ red (diagonal filling) and $|H_3|=2$ green (dot filling) nodes. Suppose that the community $c_1$ (dashed outline) with $|V(c_1)|=6$ nodes has been identified as part of the algorithm. Following the proportion of colors in the network, four of its members should be colored blue and red (two of each color), and one becomes green; without considering the extra uncolored node, $balance(c_1)=\phi=0.5$. This leaves one extra node potentially belonging to either color. 

We then consider how the balance score of the community would change if the additional node belongs in either demographic group.
If the extra node is colored green, $balance(c_1)$ increases to one, as all nodes are now equally represented in $c_1$. On the contrary, if the extra node is colored either blue or red, the balance score would decrease to $0.4$, as either of the majority classes has gained one more member in $c_1$.
The expected proportional balance score for that community, lying between those two extremes ($exp\_prop \approx 0.57$), represents the possibility that one additional node from the minority color increases community balance to a large extent, while additional nodes of any other color lead to a lower balance score.

We note that if at least two classes are equally represented in $H$, adding an extra node to an already perfectly balanced $c_1$ always decreases the expected proportional balance score slightly. This happens because, even if the extra node is a minority demographic member, another group now becomes the least represented in $c_1$. However, this has a negligible effect in practice where the community and group sizes are vastly greater, especially considering this score is meant as a rough expectation used to penalize large deviations of community balance from the group distributions.

\section{Comparison of fairness functions for real datasets}\label{app:real-data}

\begin{figure}[H]
    \centering
    \includegraphics[height=.35\textheight,keepaspectratio,angle=90]{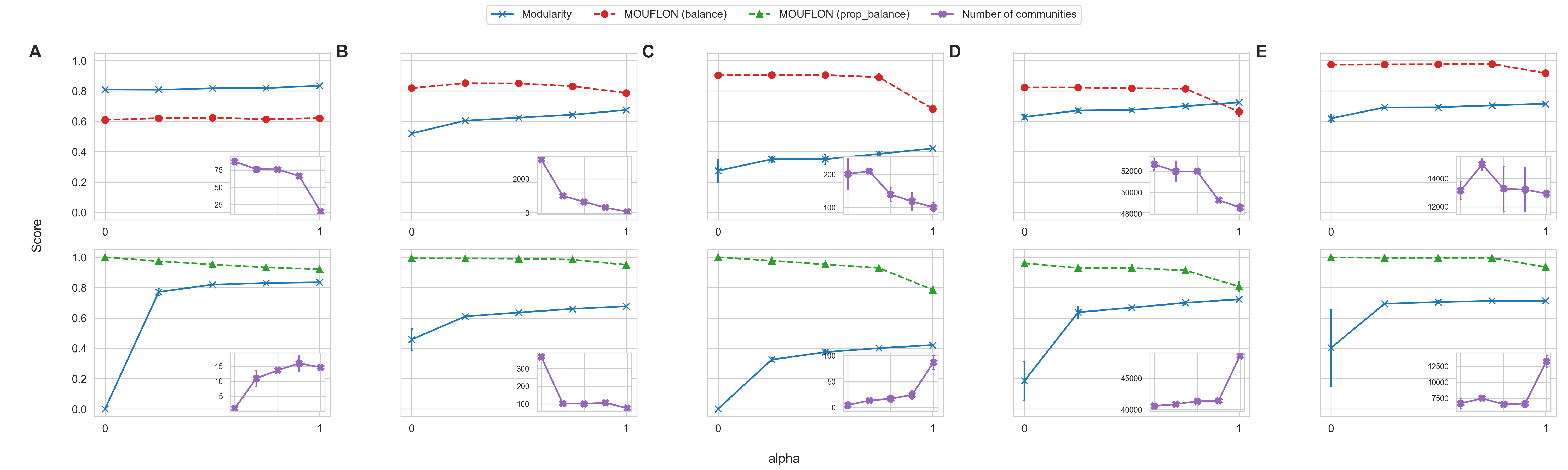}
    \caption{Comparison of average values of modularity, fairness, and number of communities (inset figures) over $\alpha$ for real-world network datasets. Communities obtained using MOUFLON (right column) methods, on (a) Facebook, (b) Deezer, (c) Twitch, (d) Pokec-a, (e) Pokec-g}
    \label{fig:real-data}
\end{figure}

In Fig. \ref{fig:real-data} we present the modularity, fairness score and number of communities obtained with MOUFLON using balance (top row) and proportional balance (bottom row) fairness, on the real-world network datasets in Table \ref{tab:real-sn}, for varying $\alpha$. 
We observe similar trends to those seen in the synthetic networks: as $\alpha$ increases, modularity tends to rise, while fairness under both metrics generally declines. However, the two fairness measures exhibit distinct behaviors. First, when fairness is measured using balance, the number of detected communities typically decreases with increasing $\alpha$, suggesting that the algorithm merges smaller communities as it prioritizes structural quality. In contrast, when using proportional balance, the opposite trend is observed in most datasets, with the exception of the Deezer network for $\alpha=0$. Second, the proportional balance scores, consistent with our synthetic data results, are more stable across $\alpha$, indicating greater robustness to the trade-off parameter.

\end{appendices}

\bibliography{references}

\end{document}